\documentclass[aip, amsmath, amssymb, reprint]{revtex4-1}
\pdfoutput=1

\usepackage{graphicx}% Include figure files
\usepackage{dcolumn}% Align table columns on decimal point
\usepackage{bm}% bold math
\usepackage{float}
\usepackage[utf8]{inputenc}
\usepackage[T1]{fontenc}
\usepackage{mathptmx}
\usepackage{etoolbox}
\usepackage[version=4]{mhchem}
\usepackage{siunitx}
\makeatletter

\def\@email#1#2{%
 \endgroup
 \patchcmd{\titleblock@produce}
  {\frontmatter@RRAPformat}
  {\frontmatter@RRAPformat{\produce@RRAP{*#1\href{mailto:#2}{#2}}}\frontmatter@RRAPformat}
  {}{}
}%
\makeatother

\newcommand{\beginsupplement}{%
        \setcounter{table}{0}
        \renewcommand{\thetable}{S\arabic{table}}%
        \setcounter{figure}{0}
        \renewcommand{\thefigure}{S\arabic{figure}}%
        \setcounter{section}{0}
        \renewcommand{\thesection}{S\arabic{section}}%
     }

\pdfoutput=1

\begin{document}

\preprint{AIP/123-QED}

\title[Characterization of Nb films for superconducting qubits using phase boundary measurements]{Characterization of Nb films for superconducting qubits using phase boundary measurements}
% Force line breaks with \\

\author{Kevin M. Ryan}
    \affiliation{Northwestern University, Department of Physics and Astronomy, Evanston, Illinois 60208, USA}
    %\email{KevinRyan2024@u.northwestern.edu}

\author{Carlos G. Torres-Castanedo}%
    \affiliation{Northwestern University, Department of Materials Science and Engineering, Evanston, Illinois 60208, USA}
\author{Dominic P. Goronzy}%
    \affiliation{Northwestern University, Department of Materials Science and Engineering, Evanston, Illinois 60208, USA}
    \affiliation{Northwestern University, International Institute for Nanotechnology, Evanston, Illinois 60208, USA}

\author{David A. Garcia Wetter}
    \affiliation{Northwestern University, Department of Materials Science and Engineering, Evanston, Illinois 60208, USA}
        
\author{Matthew J Reagor}%
    \affiliation{Rigetti Computing, Berkeley, California 94710, USA }    
\author{Mark Field}%
    \affiliation{Rigetti Computing, Berkeley, California 94710, USA }
\author{Cameron J Kopas}%
    \affiliation{Rigetti Computing, Berkeley, California 94710, USA }
\author{Jayss Marshall}%
    \affiliation{Rigetti Computing, Berkeley, California 94710, USA }
    
\author{Michael J. Bedzyk}%
    \affiliation{Northwestern University, Department of Materials Science and Engineering, Evanston, Illinois 60208, USA}
    \affiliation{Northwestern University, Department of Physics and Astronomy, Evanston, Illinois 60208, USA}
    
\author{Mark C. Hersam}%
    \affiliation{Northwestern University, Department of Materials Science and Engineering, Evanston, Illinois 60208, USA}
    \affiliation{Northwestern University,  Department of Chemistry, Evanston, Illinois 60208, USA}
    \affiliation{Northwestern University, Department of Electrical and Computer Engineering, Evanston, Illinois 60208, USA}    

\author{Venkat Chandrasekhar}
    \affiliation{Northwestern University, Department of Physics and Astronomy, Evanston, Illinois 60208, USA}
    \email{v-chandrasekhar@northwestern.edu}
    
\date{\today}% It is always \today, today,

\begin{abstract}
Continued advances in superconducting qubit performance require more detailed understandings of the many sources of decoherence. Within these devices, two-level systems arise due to defects, interfaces, and grain boundaries, and are thought to be a major source of qubit decoherence at millikelvin temperatures. In addition to Al, Nb is a commonly used metalization layer for superconducting qubits. Consequently, a significant effort is required to develop and qualify processes that mitigate defects in Nb films.  As the fabrication of complete superconducting qubits and their characterization at millikelvin temperatures is a time and resource intensive process, it is desirable to have measurement tools that can rapidly characterize the properties of films and evaluate different treatments. Here we show that measurements of the variation of the superconducting critical temperature $T_c$ with an applied external magnetic field $H$ (of the phase boundary $T_c - H$) performed with very high resolution show features that are directly correlated with the structure of the Nb films.  In combination with x-ray diffraction measurements, we show that one can even distinguish variations quality and crystal orientation of the grains in a Nb film by small but reproducible changes in the measured superconducting phase boundary.
\end{abstract}

\maketitle

Two-level systems (TLSs) in superconducting devices form significant sources of qubit decoherence at millikelvin temperatures,\cite{premkumar_microscopic_2021} and are an important factor limiting qubit performance. TLSs may exist at the interfaces between Nb films and the substrate,\cite{murthy_tof-sims_2022} in the oxide layers at the surface of Nb films,\cite{romanenko_three-dimensional_2020, verjauw_investigation_2021} at grain boundaries in Nb films,\cite{premkumar_microscopic_2021} and at defects in the Nb film itself.\cite{zarea_effects_2022} As common fabrication processes preclude the use of epitaxial single crystal Nb, a number of techniques are being explored to mitigate the effects of interfaces, grain boundaries, and native surface oxides as sources of TLSs using wafer scale processing. %For example, capping of Nb films with other metals or insulators may mitigate surface oxide formation, or annealing deposited films to increase grain size and reduce the contribution of grain boundaries.  

The ultimate test of these mitigation strategies would be to fabricate superconducting qubits with processed Nb films and measure their coherence times at millikelvin temperatures. However, this is a time consuming and expensive process which severely limits the throughput of materials development.  Alternatively, materials science characterization tools such as scanning transmission electron microscopy (STEM),\cite{altoe_localization_2020,murthy_insights_2022} secondary ion mass spectroscopy (SIMS)\cite{murthy_tof-sims_2022} or atom probe tomography (APT) can also be used to quantify the structural quality of the Nb films.  While such probes give detailed information about the disorder of films, oxides/interfaces, and concentration of impurities, the information is usually obtained for microscopic cross-sections of the device and the relationships between these observations and the superconducting properties important for qubit operation are not immediately evident.  

Conversely, macroscopic electrical transport measurements are well suited for evaluation of process changes to superconducting films.  For example, simple temperature dependent resistivity measurements can determine $T_c$, and the low temperature sheet resistance and residual resistivity ratio (RRR) provide information about the amount of disorder in the film. More sophisticated measurements can provide important microscopic parameters of the superconductivity itself.

Here, we focus on one such technique which measures the superconducting phase boundary $T_c-H$ rapidly and with high precision. From Ginzburg-Landau (GL) theory, $T_c$ of a superconducting film with an external field $H$ applied perpendicular to the plane of the film varies with $H$ as
\begin{equation}\label{eq:1}
    T_c = T_{c0} \left(1 - \frac{2 \pi \xi_0^2}{\Phi_0} H\right)
\end{equation}
where $\xi_0$ is the zero-temperature GL coherence length, $T_{c0}$ is the zero-field transition temperature, and $\Phi_0=h/2e$ is the superconducting flux quantum.\cite{tinkham_introduction_2004} Thus, $T_c$ is a linear function of $H$, and its slope combined with the measured value of $T_{c0}$ directly gives a measure of $\xi_0$. 
 
When applied to the study of arbitrary superconducting films, this analysis is most frequently performed by tracing the resistive transition of the film at various fixed fields, defining $T_c$  as $T$ at a fixed fraction of the normal state resistance, and plotting this $T_c$ as a function of $H$. This approach suffers from several drawbacks. First, in cryogenic probes used for such measurements, there is always a thermal lag between the temperature of the thermometer and that of the sample. Thus it is important to sweep slowly both up and down in temperature through the resistive transition so that the hysteresis in the traces is minimized. This is time consuming and incurs additional cryogen and operational costs. Second, the discrete nature of the magnetic fields at which $T_c$ is determined may result in features of the phase boundary being missed. To avoid these issues, we have used a feedback technique that allows for the measurement of $T_c$ as a continuous function of $H$. Our results show non-linear deviations from GL theory in the phase boundary that, via comparison with XRD measurements, appear correlated with the crystal grain structure and disorder in the films.

\section{Sample preparation and measurement}
The Nb films examined in this work were deposited via High Power Impulse Magnetron Sputtering (HiPIMS) at a nominal base pressure $\leq10^{-9}$ Torr. To study the role of the film/substrate interface 40\,nm or 155\,nm films were deposited on three different substrates:  1) intrinsically doped (001) oriented Si; 2) a-plane \ce{Al2O3} with (110) termination; and 3) c-plane \ce{Al2O3} with (006) termination. These substrates were chosen to qualify and improve upon current fabrication techniques, as (001) oriented Si is the substrate used for commercial production of many superconducting qubits,\cite{oliver_materials_2013} while \ce{Al2O3} can be used to control the epitaxial growth and orientation of Nb films\cite{wildes_growth_2001} X-ray studies show that Nb films grow as (110) oriented grains on all substrates used in this work, and show partial epitaxial growth on \ce{Al2O3} substrates (see Supp. S1). 

\begin{table}[b]
\begin{tabular}{lcccc}
Substrate                & \multicolumn{1}{l}{d {[}nm{]}} & \multicolumn{1}{l}{$T_{c0}$ [K]} & \multicolumn{1}{l}{$\Delta T_c$ [mK]} & \multicolumn{1}{l}{RRR} \\ \hline
Si (001)                   & 155                            & 9.054                           &5                                           &4.91                         \\
Si (001)                   & 40                             & 8.720                           &8                                           &3.96                         \\
% H:Si (111)                 & 155                            & 8.906                           &42                                           &3.88                         \\
% H:Si (111)                 & 40                             & 8.189                           &15                                           &2.86                         \\
a-\ce{Al2O3}           & 40                             & 8.789                           &11                                           &5.23                         \\
c-\ce{Al2O3}           & 40                             & 8.672                           &12                                           &3.29                         \\
UHV a-\ce{Al2O3}       & 40                             & 7.010$^\dagger$                 &39                                           &2.59                        
\end{tabular}
\caption{Zero field transition properties of HiPIMS deposited films presented in this work. Values are reported under feedback. RRR is derived from resistance at 300K and 10K. $^\dagger$Values reported along the $[1 \overline{1} 0]$ direction, see Fig. \ref{fig:Angular}.}
\label{tab:Film_Values}
\end{table}

\begin{figure}[t]
    \centering
    \includegraphics[width=.47\linewidth]{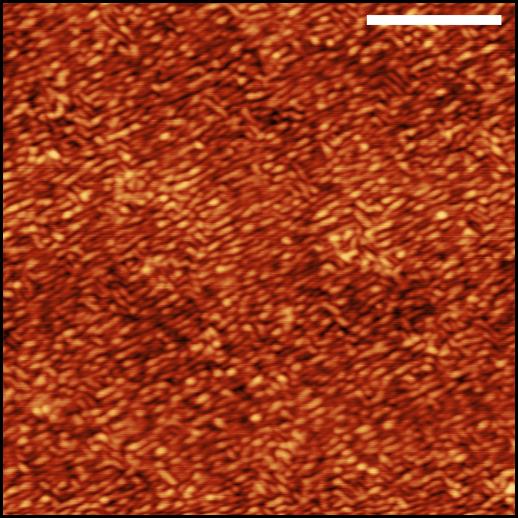}\hspace{.2cm}
     \includegraphics[width=.47\linewidth]{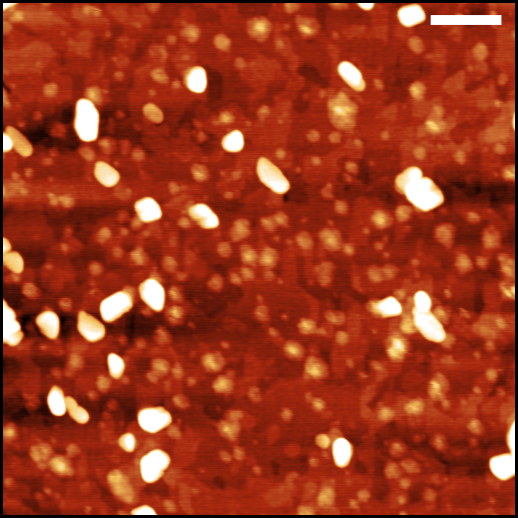}
    \caption{AFM micrographs of pristine (left) and UHV annealed (right) 40\,nm Nb films on a-\ce{Al2O3} substrates. Both scale bars are 250\,nm.}
    \label{fig:AnnealingProof}
\end{figure}

In addition, separate 40\,nm thick films deposited on a-\ce{Al2O3} substrates were vacuum annealed at 1000$^\circ$ C for 30 minutes at 10$^{-10}$ Torr.  This treatment resulted in enhanced epitaxy of the Nb, with the $[1\bar{1}0]$ Nb plane oriented along the$[\overline{1} 0 0]$ direction on the $(110)$ \ce{Al2O3} surface.  X-ray diffraction (XRD) rocking curve data show that this treatment results in the average grain size of the Nb increasing from 7\,nm to 124\,nm, with columnar growth extending from the substrates to the surface oxide (see Supp. S2).  Figure \ref{fig:AnnealingProof} shows Atomic Force Microscopy (AFM) maps of Nb films on a-\ce{Al2O3} before and after this annealing process. After deposition the films were patterned using standard photolithography techniques into \SI{1.2}{\milli\meter} by \SI{30}{\micro\meter} Hall bars, then defined using an Ar ion mill with the photoresist serving as an etch mask.   

The devices were measured in a custom probe inserted into a He dewar with a 3 T maximum field perpendicular to the plane of the films. In the probe, the devices were mounted in vacuum and thermally anchored to a variable temperature stage that is weakly coupled to the 4K flange of the probe. By controlling the power through a heater mounted on this stage, the temperature of the stage could be varied with a short response time. The accuracy of the thermometer was verified by measuring the transition of a 99.99\% pure Nb wire, which showed a $T_c$ of 9.22\,K.  The resistance of the devices were measured by custom modified Adler-Jackson ac resistance bridges,\cite{adler_system_1966} with ac excitations of $\sim100$ nA at frequencies $\leq\SI{200}{\hertz}$. Table \ref{tab:Film_Values} summarizes details of the transitions of the films included in this study. The transition temperatures of the majority of the films we measured were in the range of 8.5-9\,K with residual resistance ratios (RRR, the ratio of the room temperature resistance to $R_n$, the resistance just above the superconducting transition) of $\sim$5.  

\begin{figure}[h]
    \centering
    \includegraphics[width=\linewidth]{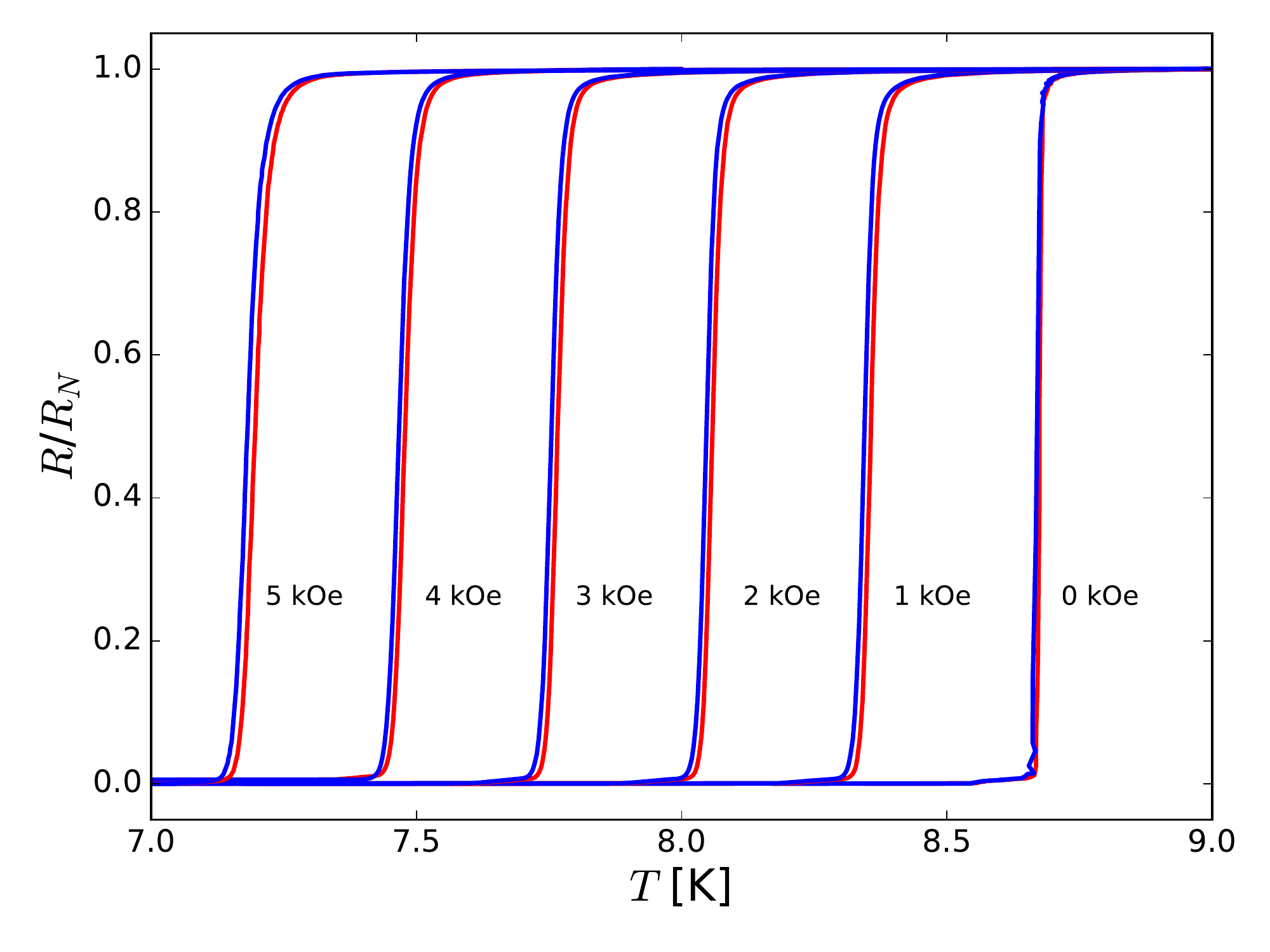}
    \caption{$R(T)$ for a 40\,nm Nb film on a a-\ce{Al2O3} substrate, normalized to the normal state resistance $R_n$ measured at various externally applied perpendicular fields obtained at a temperature sweep rate of $<5$K/hr, demonstrating $\sim$20 mK of thermal hysteresis. Red/blue curves indicate data taken during warming/cooling.}
    \label{fig:RT_Sweeps}
\end{figure}

To initially qualify the superconducting transition of the films, the power to the stage heater was ramped slowly under computer control.  Fig. \ref{fig:RT_Sweeps} shows examples of a superconducting transitions measured on a 40\,nm thick film on an a-\ce{Al2O3} substrate ramping both up and down in temperature at various magnetic fields.  For most films studied, the transition widths were $\lessapprox$ 20\,mK. Figure \ref{fig:RT_Sweeps} also illustrates the problem of using the resistive transition to accurately measure the phase boundary: even with a slow ramp rate of 5\,K/hr, the traces corresponding to sweeps up and down in temperature show hysteresis comparable to the transition width.  

To accurately measure $T_c-H$, we use a feedback technique to maintain the resistance of the sample at a specified value along the superconducting transition by adjusting the set-point resistance of the bridge to the desired value as an applied magnetic field is swept. The error signal from the bridge is then used as the input to a homemade analog proportional-integral-differential (PID) controller which controls the power to the stage heater to maintain the sample resistance. By fixing the bridge resistance at $R_n/2$, the temperature as read from the thermometer directly gives $T_c$ when the error signal is minimized by the PID under feedback. To define the transition width, one can additionally measure $T - H$ at other values of resistance, such as $5\%$ $R_n$ and $95\%$ $R_n$ to measure the transition width (see Fig. \ref{fig:HT_Al2O3_Pristine}).  This technique has been used to measure coherence lengths and anomalous Little-Parks oscillations in mesoscopic Al structures;\cite{vloeberghs_anomalous_1992} superconductivity in star shaped mesoscopic devices;\cite{dikin_nucleation_2003} and hysteretic superconducting phase boundaries arising from underlying magnetism in two-dimensional LAO/STO heterostructures;\cite{mehta_interplay_2015} but has not previously been applied to the assessment of superconducting materials for quantum applications.

\begin{figure}[h]
    \centering
    \includegraphics[width=\linewidth]{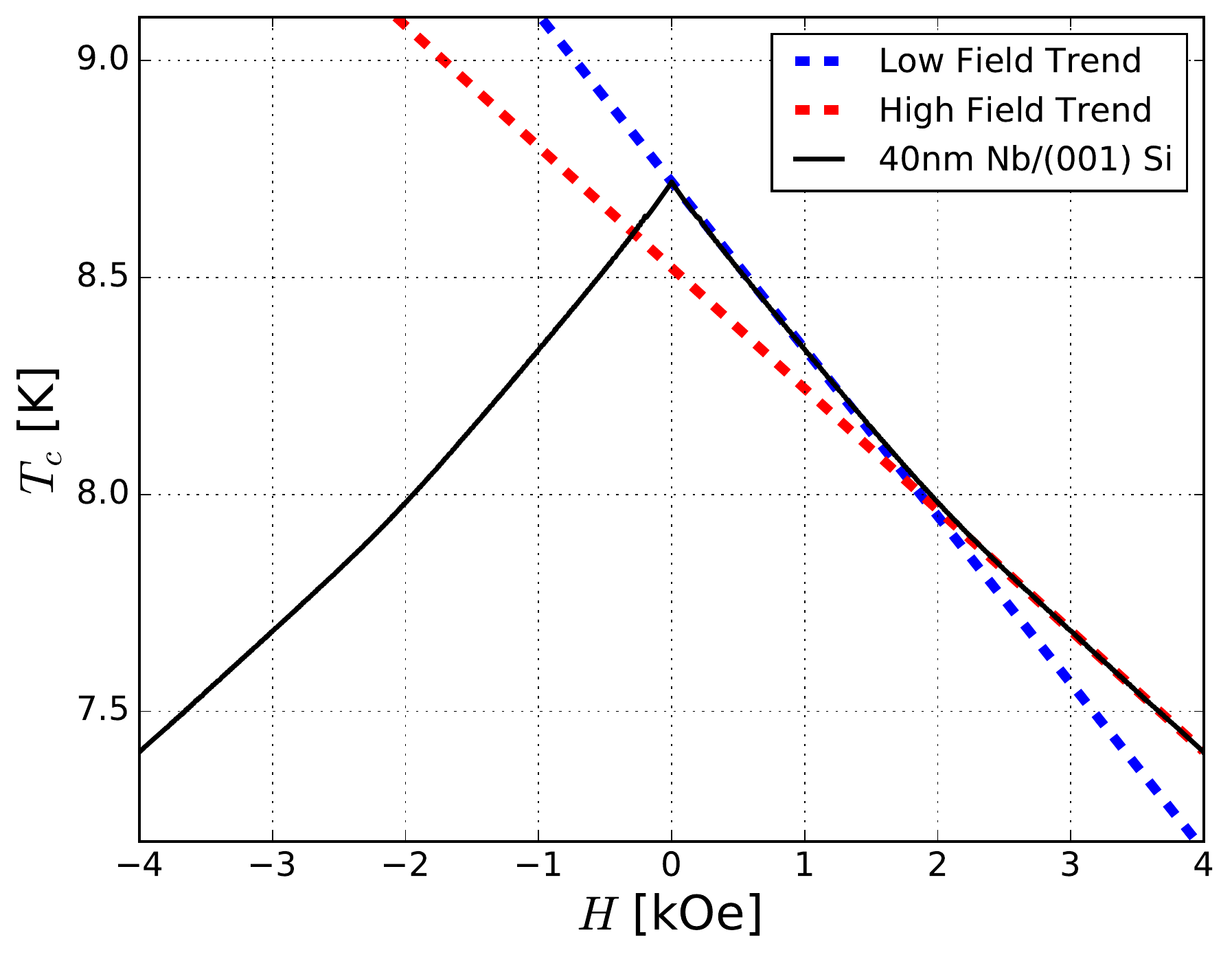}
    \caption{$T_c-H$ phase boundary measured for a 40\,nm Nb film on Si (001), taken over repeated field sweeps. The dashed lines indicate linear trends below 1 kOe (blue) and above 3 kOe (red).}
\label{fig:HT_Example}
\end{figure}

\section{Experimental results}

Figure \ref{fig:HT_Example} shows an example $T_c-H$ measurement for a 40\,nm thick Nb film on Si(001), ramping in both directions in field.  As can be seen, the curves corresponding to the different sweep directions cannot be distinguished, with the reproducibility being better than 1\,mK except at the field extrema where the sweep direction changes (see Supp. S3). As noted, GL theory predicts $T_c$ to be a linear function of $H$, with a slope determined by $\xi_0$. The traces in Fig. \ref{fig:HT_Example} show a clear nonlinear structure, with the derivative $|dT_c/dH|$ largest near zero field and decreasing asymptotically to a constant value at larger fields. This positive curvature of the superconducting phase boundary is seen to an extent in all as-deposited films measured. Before we explore the origin of this curvature, we first discuss the overall differences between Nb films grown on different substrates which can be seen by this method.

\begin{figure}[h]
    \centering
    \includegraphics[width=\linewidth]{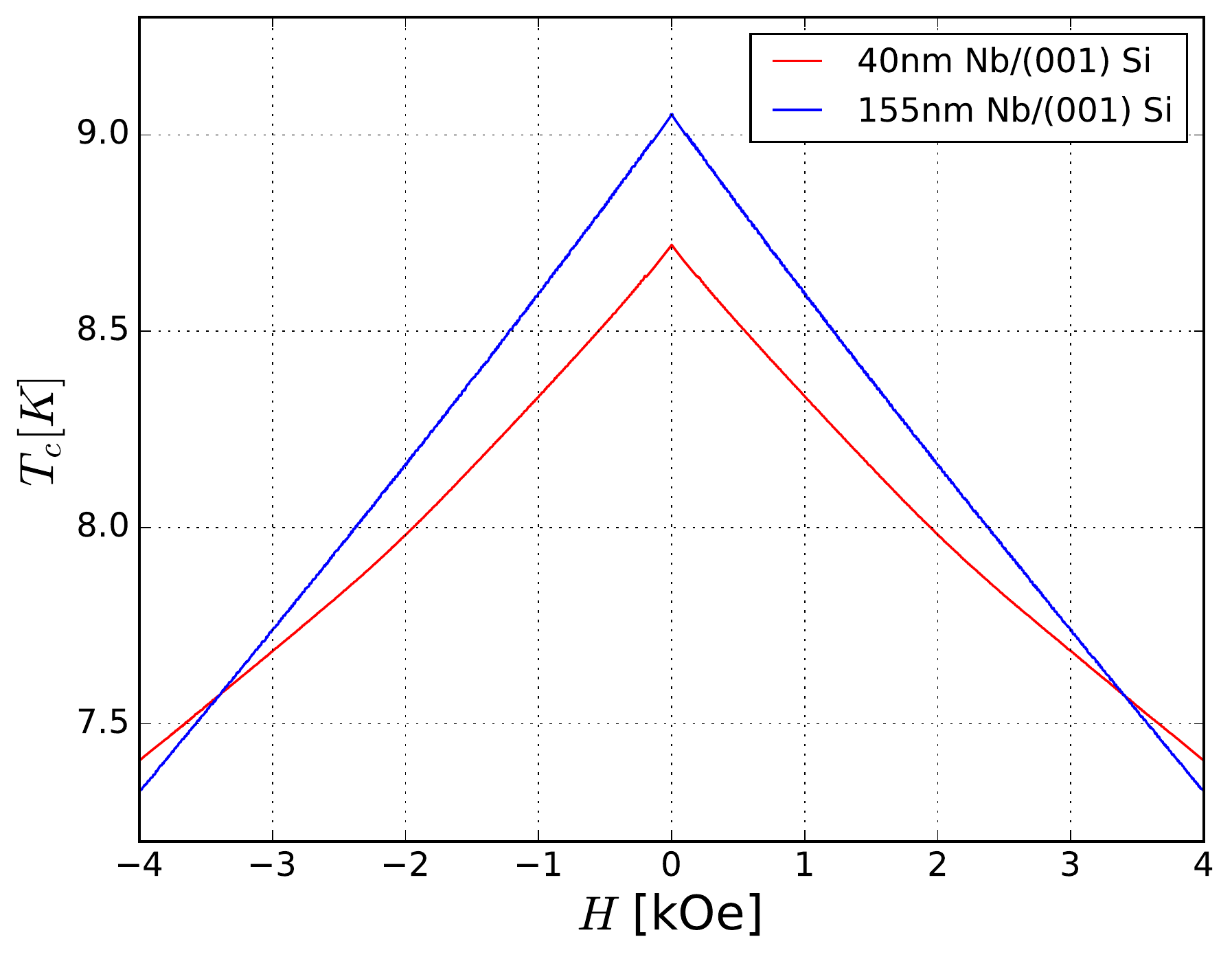}
    \caption{$T_c-H$ phase boundaries for Nb films of 40\,nm and 155\,nm thickness on Si (001). Both measurements demonstrate varying degrees of positive curvature. %Bottom: Derivatives of the raw phase boundaries, scaled to the equivalent coherence length per Eq. \ref{eq:2}.
    }
\label{fig:HT_SI_Uni}
\label{fig:HT_SI_Comp}
\end{figure}

Fig. \ref{fig:HT_SI_Comp} shows $T_c-H$ for two Nb films grown on Si (001) substrates. The $T_{c0}$ of the 40\,nm film is lower than that of the corresponding 155\,nm film, in line with existing results that show a sharp drop in $T_{c0}$'s and RRRs of Nb films below a thickness of 200\,nm.\cite{pinto_dimensional_2018} The corresponding slopes of the $T_c-H$ curves for these films give equivalent coherence lengths $\xi_0\leq$13\,nm (See Supp. S4), far less than the $\geq$30\,nm seen in bulk Nb\cite{williamson_bulk_1970} but comparable to reports for thin deposited films.\cite{pinto_dimensional_2018,bose_upper_2006} Notably, the non-linearity is more evident in the thinner 40\,nm film, appearing as a decrease in slope between 2-3\,kOe. This transition implies substantial inhomogeneity throughout the film, possibly due to greater influence of the Nb/substrate interface.   

\begin{figure}[h]
    \centering
    \includegraphics[width=\linewidth]{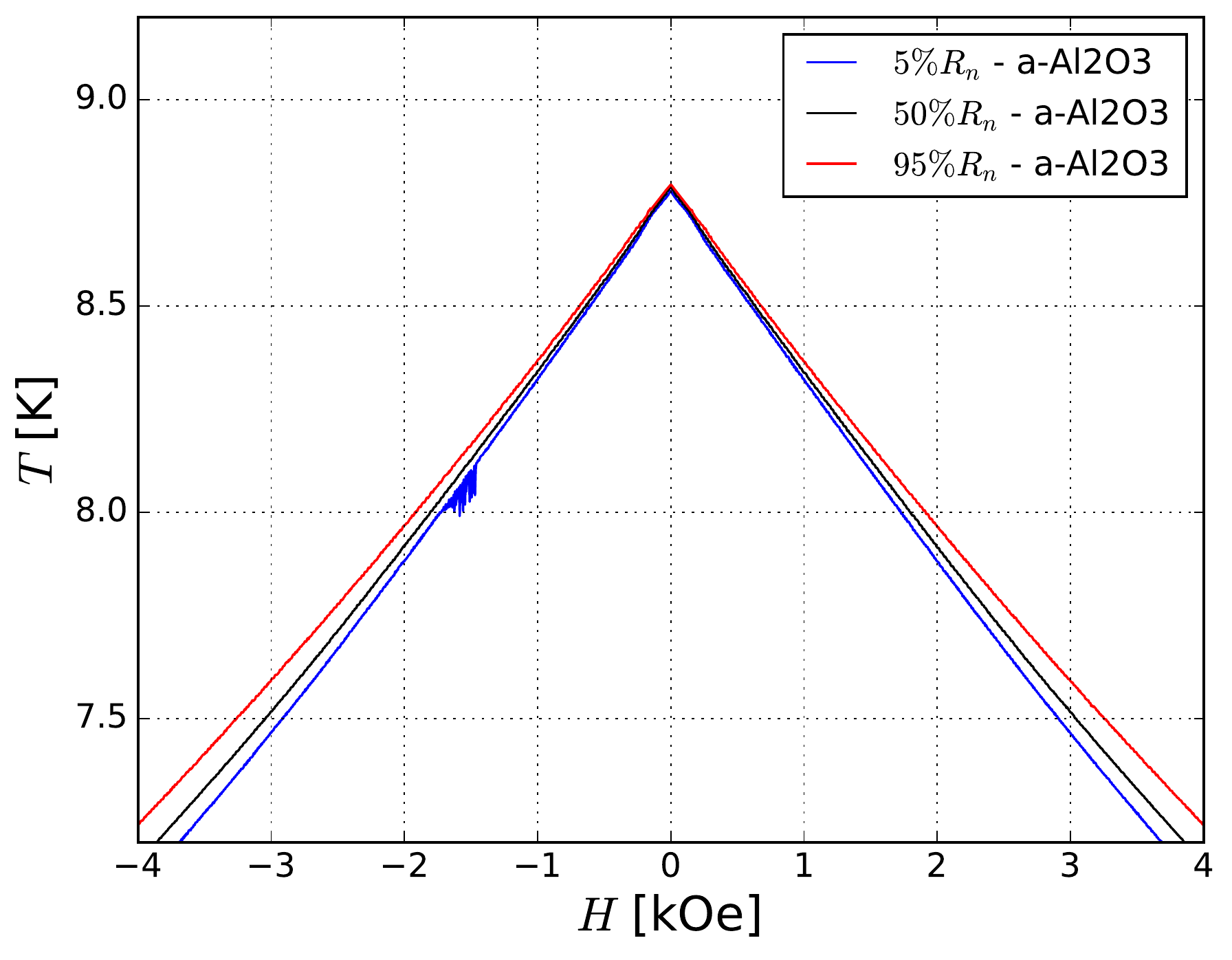}
    \includegraphics[width=\linewidth]{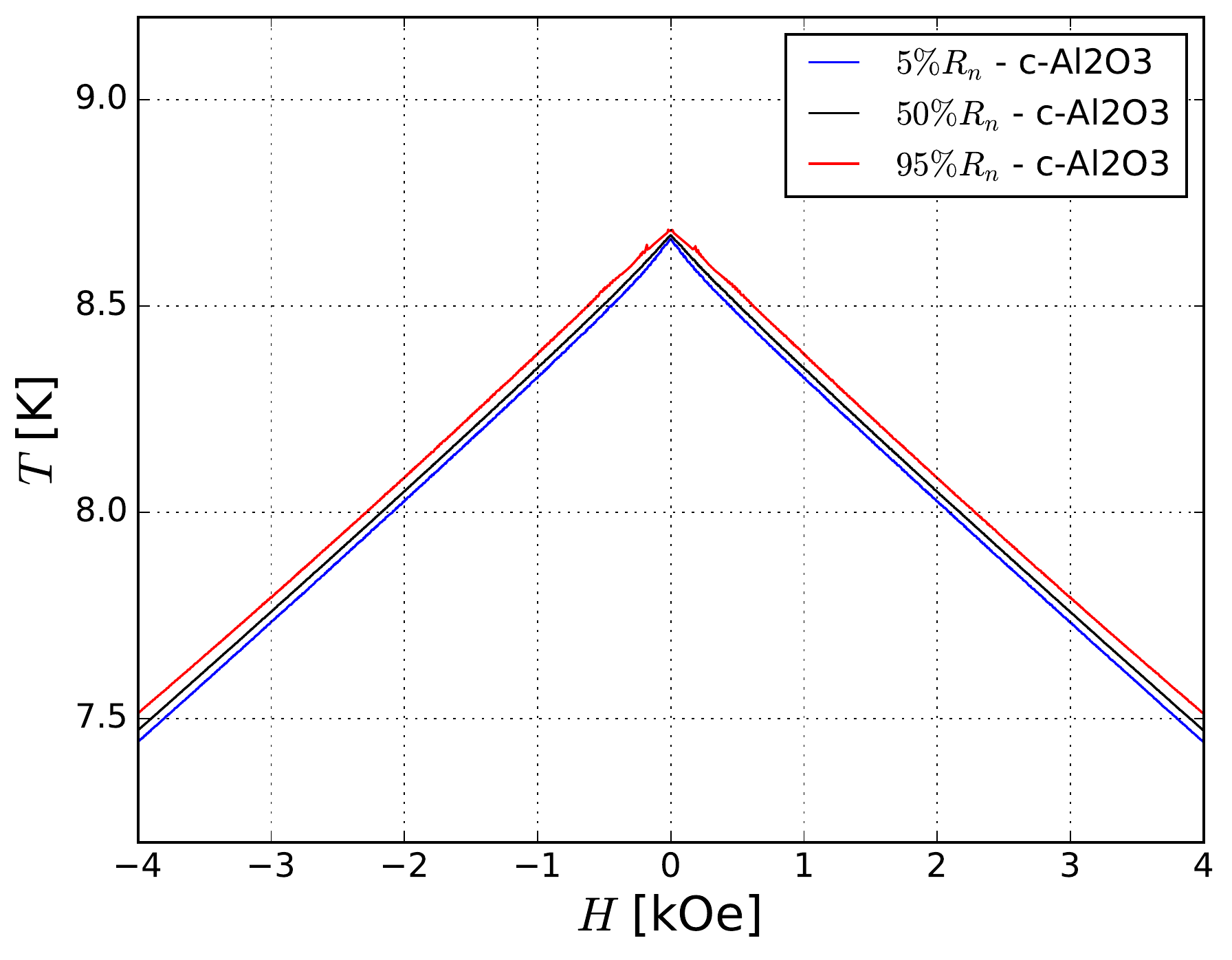}
    \caption{Temperature measured at fixed resistance values of $5\%$, $50\%$, and $95\%$ of the normal state resistance for 40\,nm Nb on a-\ce{Al2O3} (top) and c-\ce{Al2O3} (bottom) substrates. }
\label{fig:HT_Al2O3_Pristine}
\end{figure}

Fig. \ref{fig:HT_Al2O3_Pristine} shows equivalent data for 40\,nm thick Nb films grown on a- and c-plane \ce{Al2O3} substrates, where we now additionally plot the $T-H$ curves taken under feedback at 5\% and 95\% of the normal state resistance, to indicate the full width of the transition. These $T_{c0}$ values are comparable to 40\,nm Nb films grown on Si(001) (Fig. \ref{fig:HT_SI_Comp}), but a clear $T_c$ and $\xi_0$ enhancement can be seen in the a-\ce{Al2O3} film. We attribute this to improved epitaxy between the Nb(110) termination and the termination a-\ce{Al2O3} of the substrate (discussed in Supp. S1). The shape of the $T_c-H$ curves for the two substrate orientations is characteristically different, and the curves for the 5\% $R_n$ and 95\% $R_n$ diverge more at larger fields for the a-\ce{Al2O3} film than for the c-\ce{Al2O3} film. Thus there appears to be a correspondence between greater transition width broadening, positive curvature of the phase boundary, higher $T_{c0}$, and crystalline epitaxy.   
\begin{figure}[b]
    \centering
    \includegraphics[width=\linewidth]{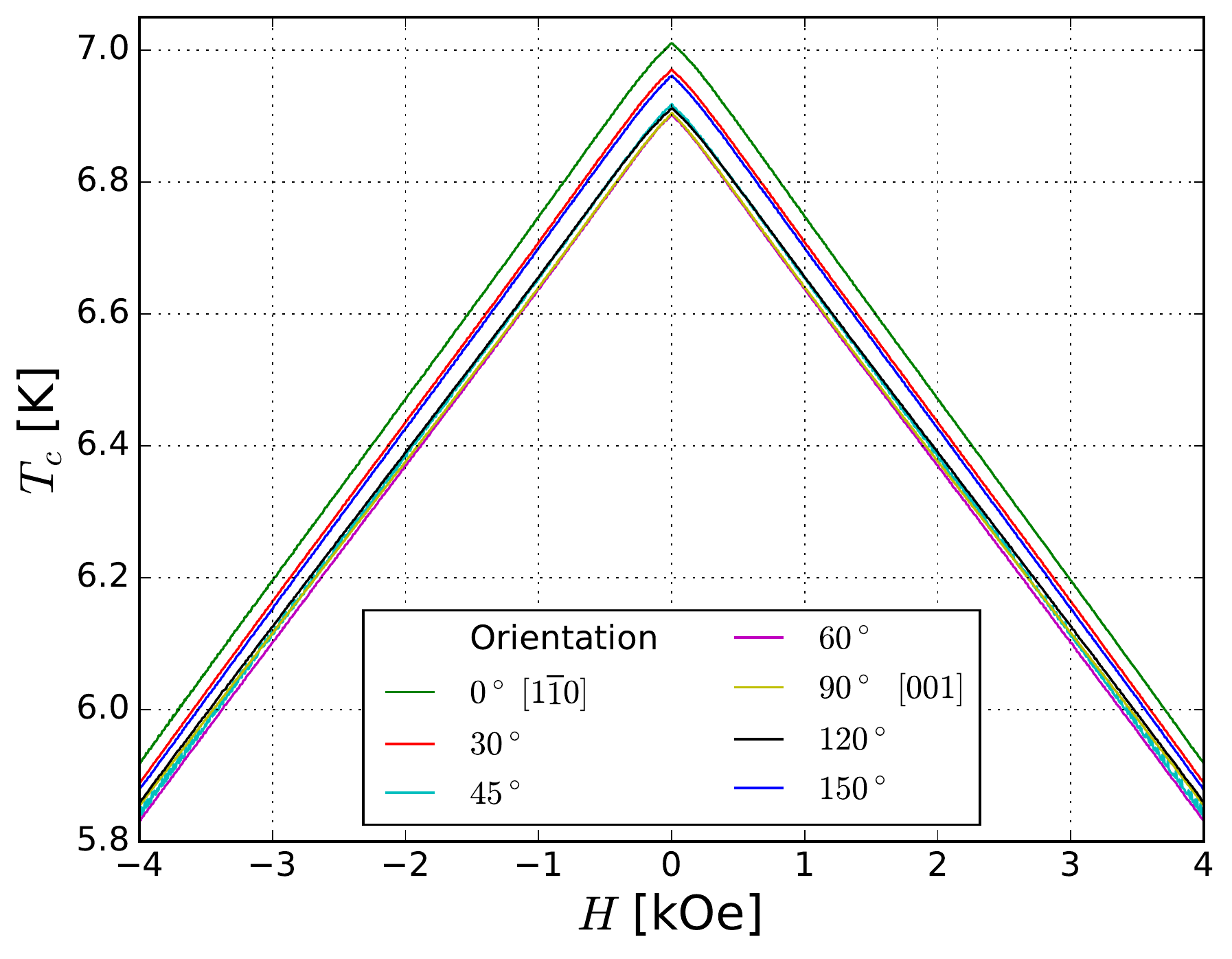}
    \includegraphics[width=\linewidth]{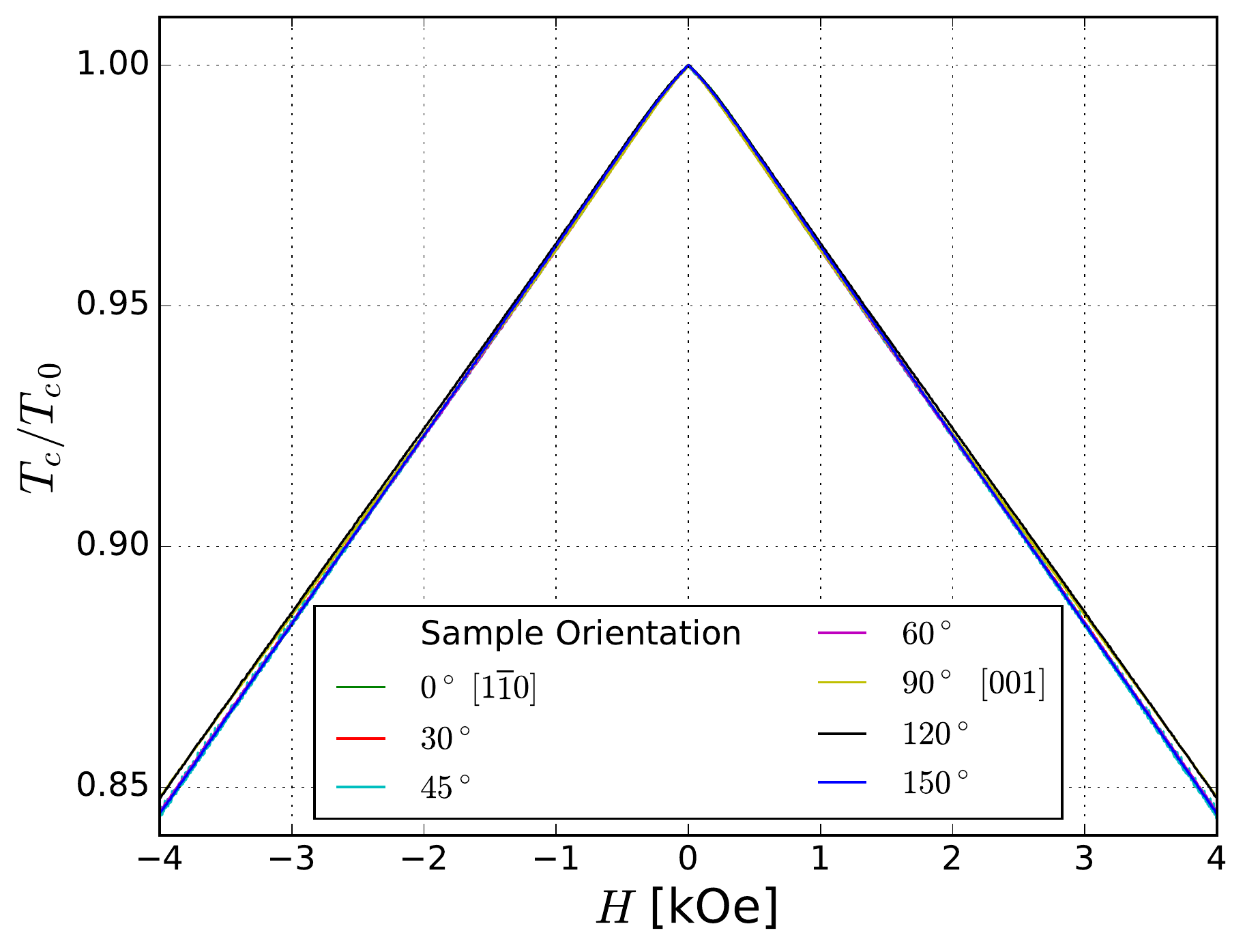}    
    \caption{Raw (top) and normalized (bottom) $T_c-H$ phase boundaries for UHV annealed 40\,nm thick Nb films on a-\ce{Al2O3} substrates.}
    \label{fig:Angular}
\end{figure}

These correlated features allow us to consider origins for the unexpected nonlinearities in $T_c-H$. As the effect appears most strongly in thinner films, and varies in magnitude for films of similar thickness on different substrates, it is likely that film granularity and inhomogeneity at the Nb-substrate interface contribute substantially. Evidence for this can be seen in our measurements of the more epitaxial UHV annealed Nb films, when studied along different crystal orientations with respect to the electronic current direction.

As noted before, Nb films grow with a (110) termination on a-\ce{Al2O3}, with the $[1\bar{1}0]$ direction of the Nb films oriented along the $[100]$ surface direction on the a-\ce{Al2O3}. XRD measurements (Supp. S1) show that UHV annealing yields columnar grains reaching from substrate to surface with lateral sizes as large at $\sim$120\,nm, much larger the $\xi_0$.  To explore the dependence of the superconducting properties as a function of the angle between the in-plane crystal direction and the current direction, Hall bars oriented with their lengths at specific angles to the Nb $[1\bar{1}0]$ orientation were fabricated from the film.  $T_c-H$ curves for these samples are shown in Fig. \ref{fig:Angular}, and are significantly different from the samples discussed earlier. First, $T_{c0}$ for these films is considerably reduced in comparison to unannealed Nb films on all substrates. However this is consistent with the reduced RRR, indicative of greater disorder in spite of the improved crystallinity, possibly due to voids generated in the film after relaxation of the crystal or diffusion of surface oxides into the bulk. Second, $T_{c0}$ appears to vary systematically with the direction of the current with respect to the crystalline axes, being a maximum when the current is aligned along the Nb $[1\bar{1}0]$ plane, and minimum close to the Nb $[001]$ direction (Fig. \ref{fig:Angular}). Third, and most striking in view of the data on Nb films on other substrates, $T_c$ is almost a perfectly linear function of $H$ except for a small field regime near $H=0$ where one now sees a \emph{negative} curvature. Normalizing these phase boundaries by $T_{c0}$, such that their slope is only defined by $\xi^2$, collapses them to a single curve (Fig. \ref{fig:Angular}(b)), implying that in spite of the variation of $T_{c0}$ with angle, $\xi_0$ is the same for all directions. Given the lack of correlation between $T_{c0}$ and $\xi$ after annealing, we believe that this can occur when the grain size is much larger than $\xi_0$.

These results enable us to propose a simple model to explain the nonlinearities in $T_c-H$ observed in polycrystalline Nb films with small, randomly oriented grains, wherein the Nb grains have a distribution of $T_{c0}$ and $\xi_0$ with higher $T_{c0}$ corresponding to longer $\xi_0$. This is possible when the GL parameter $\kappa $, the ratio of the London penetration depth $\lambda$ and $\xi_0$ increases as a function of impurity concentration, as is widely reported in Nb.\cite{narlikar_superconductivity_1966,kozhevnikov_equilibrium_2018,finnemore_superconducting_1966} Near $H=0$, the grains with the largest $T_{c0}$ contribute to $T_c-H$.  As $H$ increases, the $T_c$ of these grains is more rapidly reduced due to their longer $\xi_0$, at which point grains with smaller $T_{c0}$ and shorter $\xi_0$ dominate the transport, leading to a reduction in the slope of the $T_c-H$ curves.  For uniform epitaxial films with a single predominant crystal axis, one observes a linear behavior in $T_c-H$ as in Fig. \ref{fig:Angular}. For films with grains with two distinct crystalline axes with respect the current, one would observe two distinct slopes, as seen in Fig. \ref{fig:HT_Al2O3_Pristine}. For films with more complicated crystal grain structures, one might observe multiple slopes or even a continual change in slope at low fields. 

In summary, we observe unexpected nonlinearities in the superconducting phase boundaries of thin polycrystalline Nb films fabricated on different substrates.  These nonlinearities are due the distributions of critical temperatures and coherence lengths of crystal grains in the films.  Careful measurements of the superconducting phase boundary as demonstrated here, thus enable a rapid means of determining the quality of Nb films used in the fabrication of superconducting qubits. 

%Acknowledgment Section
This material is based upon work supported by the U.S. Department of Energy, Office of Science, National Quantum Information Science Research Centers, Superconducting Quantum Materials and Systems Center (SQMS) under Contract No. DEAC02-07CH11359. This work made use of the NUFAB facility of Northwestern University’s NUANCE Center, which has received support from the SHyNE Resource (NSF ECCS-2025633), the IIN, and Northwestern’s MRSEC program (NSF DMR-1720139). This work made use of the Jerome B. Cohen X-Ray Diffraction Facility supported by the MRSEC program of the National Science Foundation (DMR-1720139) at the Materials Research Center of Northwestern University and the Soft and Hybrid Nanotechnology Experimental (SHyNE) Resource (NSF ECCS-2025633). Additional support and equipment was provided by DURIP grant W911NF-20-1-0066. 

\bibliography{NbPhaseBoundary_Preprint_2.bib}

\cleardoublepage

\beginsupplement
\onecolumngrid

\section*{\fontsize{12}{15}\selectfont Supplementary material for: Characterization of Nb Films for Superconducting Qubits using Phase Boundary Measurements }

\section{Epitaxy of \ce{Al2O3} deposited Nb films}\label{XRD}

\begin{figure}[H]
    \centering
    \includegraphics[width=0.45\linewidth]{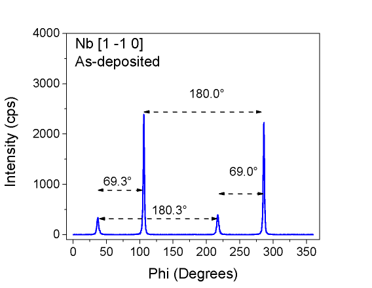} \includegraphics[width=0.45\linewidth]{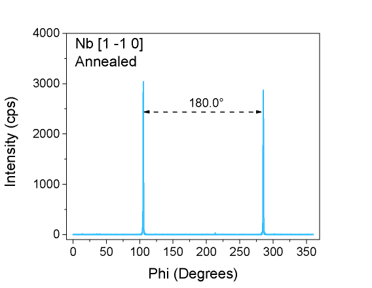}
    \caption{In-plane XRD of the Nb($110$)/\ce{Al2O3}($110$) film before (left) and after the UHV annealing (right) at $1000^\circ$C for 30 minutes. }
    \label{XRRFig}
\end{figure}

Epitaxial growth of Nb on \ce{Al2O3} was confirmed via X-ray characterization using a Rigaku Smartlab 9kW diffractometer. Specular X-ray diffraction and X-ray reflectivity (XRR) measurements were performed at $\lambda=\SI{1.5406}{\angstrom}$ (Cu K$\alpha_1$) with a Ge (220)x2 monochromator. In the case of in-plane X-ray diffraction measurements, an incident and receiving $0.5^\circ$ Soller slits were utilized with no monochromator (Cu K$\alpha$).

Epitaxy of the \ce{Al2O3}(110) grown Nb films was studied by X-ray diffraction before and after the UHV annealing (Fig. \ref{XRRFig}). In-plane XRD measurements show 2-fold symmetry in the Nb [$1 \overline{1} 0$] in-plane Bragg peak (Fig. \ref{XRRFig}) for both the as-deposited and UHV-annealed samples. However, an additional 2-fold symmetry domain is observed in the as-deposited film suggesting an impartial epitaxy for the Nb[$1\overline{1}0$] || Al2O3[$1 \overline{1} 0 4]$ orientation. The UHV annealed sample shows an overall improvement in the epitaxial orientation of the films, with two sharp Nb [$1 \overline{1} 0$] peaks, 180° apart, as expected for a uniform epitaxially aligned film.

\section{X-ray properties of UHV annealed Nb on \ce{Al2O3} (110)}\label{Annealing}
 \begin{figure}[H]
    \centering
    \includegraphics[width=0.5\linewidth]{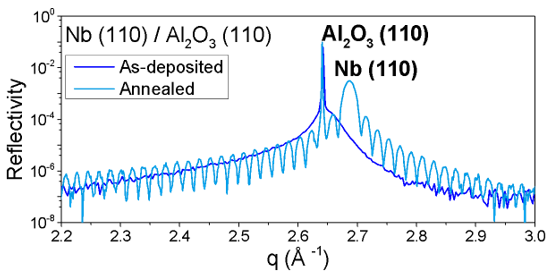}
    \caption{Specular XRD of the Nb(110)/Al2O3(110) film before and after the UHV annealing at 1000°C for 30 min. }
    \label{SXRDFig}
\end{figure}

The ultra-high vacuum (UHV) annealing was performed in a custom-built chamber with a base pressure of $\approx\SI{2E-10}{Torr}$. The annealing was monitored with a pyrometer. The final temperature of $1000^\circ\,$C was held for 30 minutes at a base pressure of $\approx\SI{2E-9}{Torr}$. UHV conditions were necessary to increase the crystallinity of the films without oxidizing the Nb film. The morphology of the film was analyzed by atomic force microscopy in an Asylum Cypher (Oxford Instruments) using tapping mode, and is shown in Fig. \ref{fig:AnnealingProof} of the main text. 

\begin{table}[H]\centering
\begin{tabular}{lcc}
Nb/a-\ce{Al2O3} (110) & \multicolumn{1}{l}{Pristine} & \multicolumn{1}{l}{UHV Annealed} \\ \hline
Nb (110)              & \SI{2.649}{\angstrom^{-1}}     & \SI{2.687}{\angstrom^{-1}}         \\
$a$                   & \SI{3.354}{\angstrom}        & \SI{3.307}{\angstrom}            \\
Lattice Expansion             & 1.45\%                       & 0.02\%                           \\
$L_z$                 & -                            & \SI{327}{\angstrom}              \\
$L_x$                 & -                            & \SI{867}{\angstrom}              \\
$d$                   & \SI{320}{\angstrom}          & \SI{333}{\angstrom}              \\
Nb Roughness          & \SI{7}{\angstrom}            & \SI{5}{\angstrom}                
\end{tabular}
\caption{XRD/XRR comparison of pristine and UHV annealed Nb films on a-\ce{Al2O3}, where $a$ is the BCC lattice parameter, $L_z$ is the measured vertical grain size, $L_x$ is the measured horizontal grain size, and $d$ is the measured thickness of the Nb layer.}\label{XRDTable}
\end{table}

The as-deposited film shows a wide Bragg Nb ($110$) peak suggesting low crystallinity and a large offset with the Nb(110) bulk value as seen in Fig. \ref{SXRDFig}. After annealing, the peak shifts to the expected bulk Nb position. The sharpening of the Nb (110) Bragg peak and the appearance of thickness fringes is evidence of improved crystallinity of the film after annealing. Table \ref{XRDTable} summarizes these findings. The vertical grain size of the annealed film extracted from XRD rocking curve measurements reaches a value close to the predicted thickness by XRR. This suggests that the grains form a columnar structure up to the surface of the film (~35\,nm), which is ideal for epitaxially grown Nb. While the as-received sample has a contraction in the lattice (-1.28\%), expected from the expansion in the specular directions, while the annealed sample shows a slight expansion (0.17\%), which may have created voids across the film after relaxation.

\section{Phase boundary reproducibility and errors}\label{RepoError}

\begin{figure}[H]
    \centering
    \includegraphics[width=0.45\linewidth]{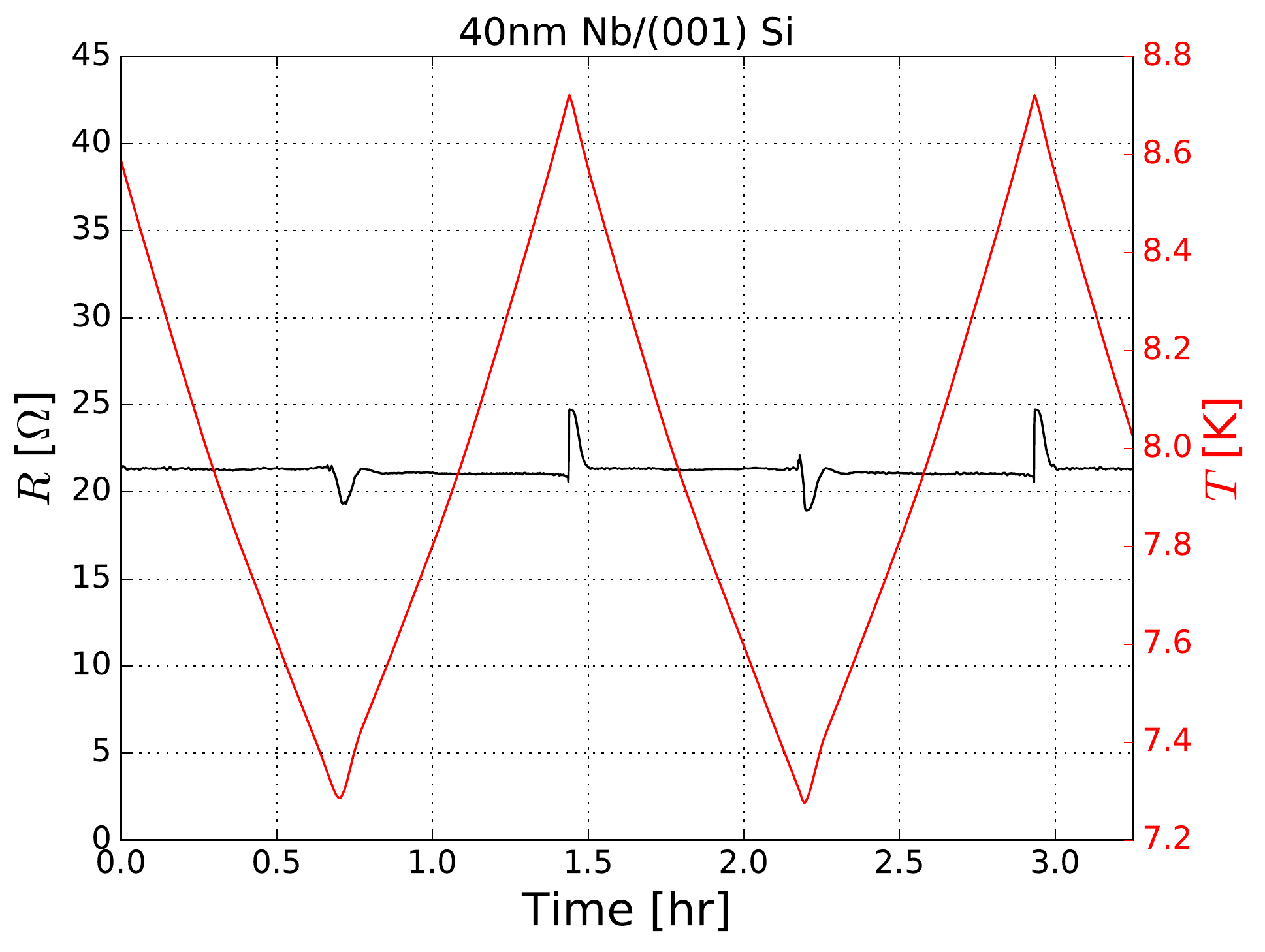}    \includegraphics[width=0.45\linewidth]{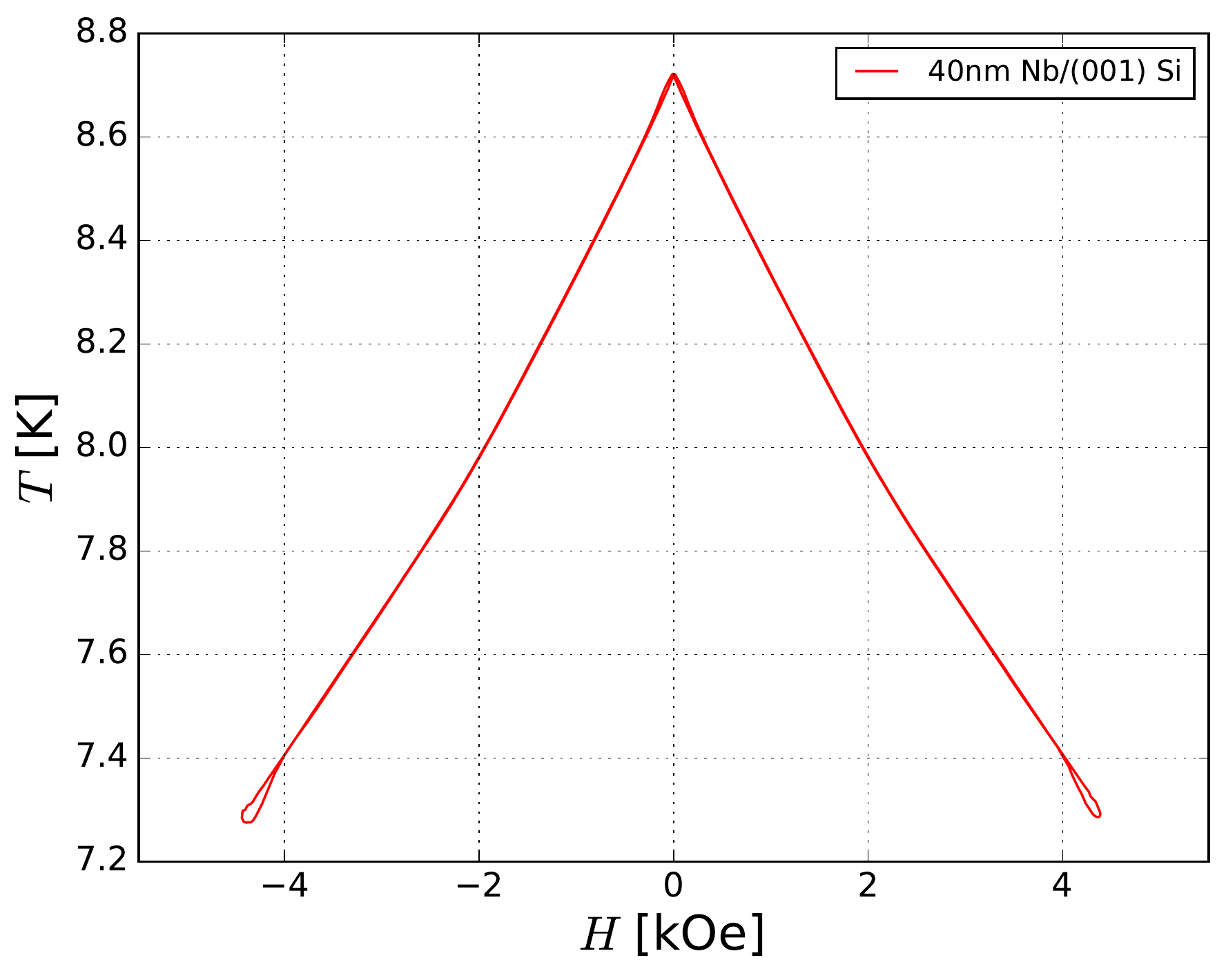}
    \caption{Left: Measured $R$ and $T_c$ for a 40\,nm Nb film on (001) Si over two full traces over the field range presented in Fig. \ref{fig:HT_SI_Comp}. Right: Raw $T-H$ data for the same 40\,nm Nb film \ref{fig:HT_SI_Comp}}
    \label{fig:Time}
\end{figure}

With sufficient tuning of the PID controller, typically the resistance of the sample can be fixed with mK precision at fixed field. As the midpoint of the transition changes under applied fields, small resistance offsets on the order of 1\% of the normal state resistance can occur as the sample is heated or cooled. This ``thermal lag'', for samples with sharp transitions, yields negligible $T_c$ hysteresis on the scale of the change in $T_c$ in field. This can be seen by examining raw time series data presented in Fig. \ref{fig:Time}

The more prominent regions where error occours are at the field extrema and at $H=0$, where the time rate of change $dT_c/dt$ changes sign, causing a rapid change in the error signal provided to the PID controller, which requires a finite integration time to return to the midpoint of the transition. By monitoring the sample resistance/error signal during a phase boundary measurement, one obtains a direct detection for periods where the sample temperature is not at $T_c$ at these extrema. For the results presented in the main text, we have suppressed these errors in two ways. First, by sweeping past the field regime of interest (typically by 500\,Oe or more), the field reversal error can be safely ignored as the PID controller brings the sample back to $T_c$ during this overshoot. Second, for errors near $H=0$, data-points which have large error signals can be excluded by a cutoff criteria. Again, for the results presented in the main text, data points where the resistance/error signal exceeds $3\sigma$ have been filtered out before interpolation/plotting. By definition, these points cannot be measurements of $T_c$ under the $50\%$ $R_n$ criteria. Raw data, without filtering or supression of the field range for a 40\,nm film is also shown in Fig. \ref{fig:Time}.

\section{Effective Coherence Length}\label{Ecoh}
\begin{figure}[H]
    \centering
    \includegraphics[width=0.5\linewidth]{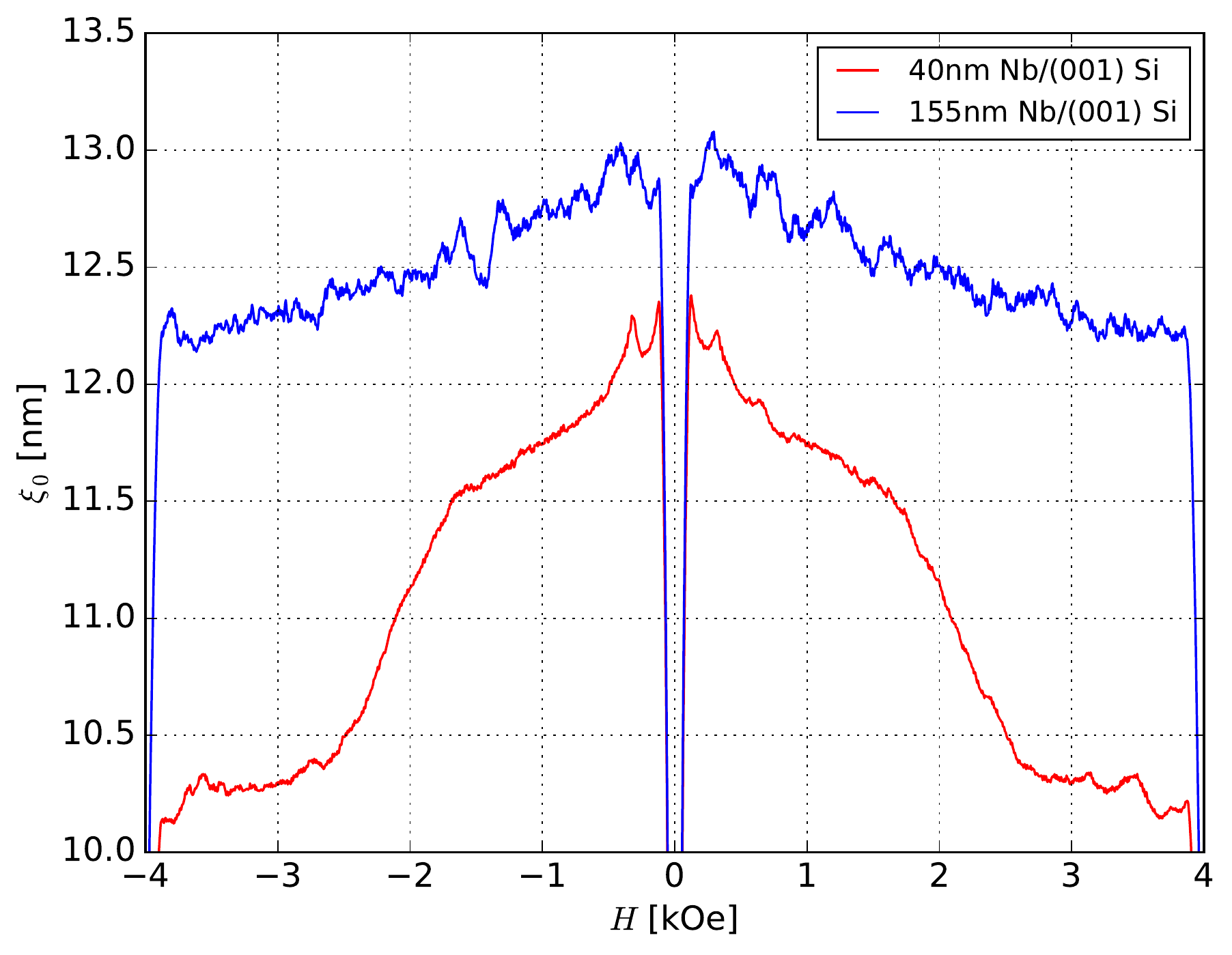}
    \caption{Effective coherence length for Nb films on (001) Si substrates.}
    \label{fig:Coherence001}
\end{figure}

For the examination of real phase boundary data, one may chose to redefine $\xi_0$ in terms of the derivative of $T_{c0}(H)$ from Eq. \ref{eq:1} as
\begin{equation}\label{eq:2}
    \xi_0 \equiv \sqrt{\left|\frac{dT_{c}}{dH} \frac{\Phi_0}{2\pi T_{c0}}\right|}
\end{equation}

In the case of small non-linearity, this treatment can reveal subtle curvature of the dataset, which can then be treated in terms of variation of the effective GL coherence length. Considering the data presented in Fig. \ref{fig:HT_SI_Comp}, when plotted in this form, a sharp inflection in the curvature of the phase boundary can be seen to be associated with a $\approx 1$\,nm decrease the in effective coherence length for the 40\,nm Nb film on (001) Si.

Numerical differentiation was performed with a second order Savitzky-Golay filter, with a window size of 155\,Oe, which causes the dip at zero field where ${dT_{c}}/{dH}$ reverses sign, and at the extrema of the data set where the range is restricted.

\clearpage

\end{document}